\documentclass[twocolumn,tighten]{aastex61}
\usepackage{amsmath}

\shorttitle{Misaligned Accretion and Jet Production}
\shortauthors{King \& Nixon}

\begin{document}

\title{Misaligned Accretion and Jet Production}

\correspondingauthor{Andrew King}
\email{ark@leicester.ac.uk}

\author[0000-0002-2315-8228]{Andrew King}
\affiliation{Theoretical Astrophysics Group, Department of Physics and Astronomy, University of Leicester, Leicester, LE1 7RH, UK}
\affiliation{Astronomical Institute Anton Pannekoek, University of Amsterdam, Science Park 904, NL-1098 XH Amsterdam, Netherlands}
\affiliation{Leiden Observatory, Leiden University, Niels Bohrweg 2, NL-2333 CA Leiden, Netherlands}

\author[0000-0002-2137-4146]{Chris Nixon}
\affiliation{Theoretical Astrophysics Group, Department of Physics and Astronomy, University of Leicester, Leicester, LE1 7RH, UK}

\begin{abstract}
Disc accretion on to a black hole is often misaligned from its spin axis. If the disc maintains a significant magnetic field normal to its local plane, we show that dipole radiation from Lense--Thirring precessing disc annuli can extract a significant fraction of the accretion energy, sharply peaked towards small disc radii $R$ (as $R^{-17/2}$ for fields with constant equipartition ratio). This low--frequency emission is immediately absorbed by surrounding matter or refracted towards the regions of lowest density. The resultant mechanical pressure, dipole angular pattern, and much lower matter density towards the rotational poles create a strong tendency to drive jets along the black hole spin axis, similar to the spin--axis jets of radio pulsars, also strong dipole emitters. The coherent primary emission  may explain the high brightness temperatures seen in jets.  The intrinsic disc emission is modulated at Lense--Thirring  frequencies near the inner edge, providing a physical mechanism for low--frequency QPOs. 

Dipole emission requires nonzero hole spin, but uses only disc accretion energy. No spin energy is extracted, unlike  the Blandford--Znajek process. MHD/GRMHD formulations do not directly give radiation fields, but can be checked post--process for dipole emission and so self--consistency, given sufficient resolution. Jets driven by dipole radiation should be more common in AGN than in X--ray binaries, and in low accretion rate states than high, agreeing with observation. In non--black--hole accretion, misaligned disc annuli precess because of the accretor's mass quadrupole moment, similarly producing jets and QPOs.
\end{abstract}

\keywords{astrophysical jets --- quasiperiodic oscillations --- accretion discs --- black hole physics\vspace{0.5in}}

\section{Introduction} \label{sec:intro}

Accretion on to a black hole is the most efficient way of extracting energy from ordinary matter. Because the matter generally has angular momentum, accretion almost always occurs through a disc. Until fairly recently most discussions of disc accretion assumed complete axisymmetry, and so implicitly a disc axis aligned with the black hole spin. But it has become clear that the opposite assumption -- disc and spin misaligned -- may well be more likely. This is natural for supermassive black holes in active galactic nuclei (AGN) as there is no obvious preferred plane for accretion \citep{King:2005aa,King:2006aa}. But even in close stellar--mass binary systems, where accretion is usually in the binary orbital plane, it is reasonable to expect some misalignment in almost all cases. This is particularly so if the black hole was born in a supernova rather than some form of quiet collapse. A supernova kick is likely to give an initial spin misalignment from the binary axis, and the asymmetry is likely to persist despite mass accretion from the companion star \citep{King:2016aa}. Even without a supernova,  completely aligned disc accretion requires tides to have made the spins of the two stars fully parallel before formation of the black hole. In this sense, the assumption of completely axisymmetric accretion may be a singular limit, ruling out various significant effects.

The frequent misalignment of accreting black hole systems carries an important implication, because of Hawking's theorem that any stationary state involving a black hole and external fields must be either static (i.e nonrotating) or axisymmetric \citep{Hawking:1972aa}. So a spinning black hole in a non--axisymmetric situation must either try to lose all its spin energy or evolve towards axisymmetry. In the second case the hole and the surrounding matter and fields must experience torques trying to remove the asymmetry. 

For example, misaligned disc orbits experience differential \cite{Lense:1918aa} precession, so that viscous torques dissipate energy as neighboring disc annuli interact \citep{Bardeen:1975aa,Nixon:2016aa}, causing the disc plane to warp. This can align the central disc region with the hole's spin plane while the outer disc remains misaligned, if fed mass maintaining this state, e.g. from a binary companion. Other effects such as disc breaking (a near--discontinuous change of disc inclination; \citealt{Nixon:2012aa}) and tearing (disc annuli precess independently; \citealt{Nixon:2012ad,Dogan:2015aa,Nealon:2015aa,Nealon:2016aa}) can also appear. 

In this paper we consider the effects of misaligned magnetic fields on black holes. For a black hole immersed in a {\it fixed} external magnetic field (i.e. one with sources of far greater inertia than the hole) the result is already known. \citet{King:1977aa} calculated the systemic torque explicitly by solving the Einstein--Maxwell equations in a Kerr background, finding a Newtonian torque
\begin{equation}
  {\bf T} = \frac{2G^2}{3c^5}M({\bf J}\times{\bf B})\times{\bf B}
  \label{T}
\end{equation}
acting on the hole--field system. ($M$ is the hole mass, ${\bf B}$ the magnetic field at infinity and ${\bf J}$ the hole's angular momentum). Since the sources of the field cannot move, this shows that the hole aligns its spin with the field by suppressing the misaligned angular momentum component $J_{\perp}$ exponentially on a timescale 
\begin{equation}
  t_h = \frac{3c^5}{2G^2MB^2},
  \label{th}
\end{equation}
The form of (\ref{T}) means that there is no precession as the hole aligns. The parallel angular momentum component $J_{\|}$ remains fixed, so that the total angular momentum $J = |{\bf J}|$ decreases on the timescale (\ref{th}), extracting rotational energy from the hole, and therefore reducing the mass $M$ to prevent the area of the event horizon decreasing. \citet{King:1977aa} used their expression (\ref{T}) to confirm that the effect of typical interstellar magnetic fields in aligning black hole spins is completely negligible\footnote{Forty years after its first appearance, this original version of magnetic alignment -- a black hole lining up with an external field with much greater inertia -- has still not found an astrophysical application. The only conceivable one appears to involve black--hole formation, perhaps in gamma--ray bursts. A huge burst of luminosity might occur if a collapsing stellar core has a spin misaligned from a massive envelope anchoring a strong magnetic field \citep[cf][]{Kim:2003aa}. Given the labour involved in the calculation in \cite{King:1977aa} (the first author's handwritten algebra covers 65 pages of large--format computer printout paper) an application would be welcome.}.

Although it was not noted at the time, the form of (\ref{T}) is highly significant. With the substitutions $\boldsymbol{\mu} = R_g^3\boldsymbol{B},\, \boldsymbol{\omega} = \boldsymbol{J}/MR_g^2$ (so $\omega = \left|\boldsymbol{\omega}\right| = ac/R_g$) where $R_g = GM/c^2$ is the gravitational radius, we get
\begin{equation}
  \boldsymbol{T} = \frac{2\omega^2}{3c^3}(\boldsymbol{\omega}\times\boldsymbol{\mu})\times\boldsymbol{\mu}
  \label{Td}
\end{equation}
which is the formula for the effective reaction torque produced by radiation from a magnetic dipole of moment $\boldsymbol{\mu}$ rotating with angular velocity $\boldsymbol{\omega}$ \citep[e.g.][]{Davis:1970aa}. The effect of immersing the black hole in a fixed misaligned field is evidently equivalent to inducing an effective dipole moment within its ergosphere. This is forced to rotate, and so 
radiates away the misaligned black hole spin component via circularly--polarized dipole emission along the spin axis.

\section{Misaligned Magnetic Disc Accretion}
A more realistic case of misaligned magnetic fields occurs if the fields are anchored orthogonal to the local planes of an accretion disc which is itself misaligned. This is a natural assumption, given that magnetic fields are generally thought to be fundamental to driving accretion through a disc, via the magnetorotational instability. Unless the fields are somehow completely contained within the disc, disc annuli have an effective dipole moment. The fixed magnetic field case considered in the Introduction then corresponds to the limiting (but unlikely) case of a disc with far greater inertia than the hole. In any realistic case, the disc annuli themselves must precess around the black hole spin under the Lense--Thirring (LT) effect. Since they carry an orthogonal magnetic field component, and the precessional motion is around the black hole spin axis, this produces electromagnetic dipole emission with a sine--squared angular pattern axisymmetric about the instantaneous acceleration direction. As this centrifugal acceleration itself rotates about the black hole spin axis at the LT frequency, we get time--averaged angular dependence
\begin{equation}
  \frac{{\rm d}L_{\rm dip}}{{\rm d}\Omega} = \frac{\omega^4\mu^2\sin^2\beta}{8\pi c^3}(1 + \cos^2\theta)
  \label{dP/do}
\end{equation}
where $\mu = BR^3$ is the magnetic moment ($B$ is the poloidal field and $R$ the radius of the ring, with $r = R/R_{\rm g}$, $\theta$ the polar angle measured from the black hole spin axis, and $\beta$ is the angle between the disc normal and this axis. (This step is directly analogous to going from the Thomson scattering cross--section for linearly polarized to unpolarized light, cf e.g. \citealt{Rybicki:1979aa}). The total time--averaged power is
\begin{equation}
  L_{\rm dip} = \frac{2\omega^4\mu^2\sin^2\beta}{3c^3}
  \label{P}
\end{equation}
emitted coherently at the circular precession frequency 
\begin{equation}
  \omega = \omega_{\rm LT} \simeq \frac{2ac}{R_gr^3}.
  \label{LT}
\end{equation}
This simple dependence is almost exact at disc radii larger than the ISCO, but becomes much more complex close to this radius \citep[cf.][]{Wilkins:1972aa,Motta:2018aa}. The approximation is adequate for the purposes of this paper. Together these give 
\begin{equation}
  L_{\rm dip} = \frac{128}{9\alpha}\left(\frac{R}{H}\right)\left(\frac{v_A}{c_s}\right)^2\frac{\dot Mc^2}{r^{17/2}}\left(1 - \frac{x^{1/2}}{r^{1/2}}\right)a^4\sin^2\beta
  \label{Ld}
\end{equation}
where $R_{\rm isco} = x R_{\rm g}$, $a = (x^{1/2}/3)[4-(3x-2)^{1/2}]$ \citep{Bardeen:1970aa}, and we have used the equations \citep[e.g.][]{Frank:2002aa} for a thin accretion disc with accretion rate $\dot M$, viscosity parameter $\alpha$, local scaleheight $H$, sound speed $c_s$ and Alfv\'en velocity $v_A$. The very steep $r^{-17/2}$ radial dependence means that the dipole emission from a disc is dominated by that emitted from the innermost radius where any significant inclination remains. Given that
\begin{equation}
  \frac{L_{\rm dip}}{L_{\rm acc}} = \frac{128}{9\eta\alpha}\left(\frac{R}{H}\right)\left(\frac{v_A}{c_s}\right)^2 \frac{a^4\sin^2\beta}{r^{15/2}}\left(1 - \frac{x^{1/2}}{r^{1/2}}\right)
  \label{ll}
\end{equation}
where $L_{\rm acc}(R) \simeq \eta\dot M c^2r^{-1}$, with $\eta \sim 0.1$, is the local accretion luminosity (strictly, gravitational binding energy release) we can see that even small disc inclinations $\beta$ can give significant dipole emission for poloidal magnetic fields not too far from equipartition ($v_A \sim c_s$).

Equation (\ref{Ld}) suggests that the dipole emission, and perhaps therefore the jet power, may have a strong dependence on the black hole parameter $a$. But as we mentioned above, we have not yet investigated the sufficient global alignment conditions for jet production. These may well restrict the likely range of $a$, so this point remains unclear.

Since the precession does no work on the black hole spin, the energy loss (\ref{P}) must come from the accretion energy released as the annulus contracts. The associated angular momentum loss is carried off by the circularly polarized emission along the black--hole spin axis, which produces a torque of the form (\ref{Td}). As this implies that ${\rm d {\bf \omega}/{\rm d}t}$ is orthogonal to ${\bf \mu}$, only the misaligned component of ${\bf \omega}$ is reduced. The net result of dipole emission is that the annulus loses gravitational energy and tends to align or counteralign with the black hole spin (i.e.  $\sin^2\beta$ decreases) on the dipole radiation timescale. Contraction of the annulus must involve dissipation, so $L_{\rm dip}$ cannot exceed some fraction of $L_{\rm acc}$. The very steep $r$--dependence of (\ref{ll}) means that $L_{\rm dip} \sim L_{\rm acc}$ for $r \sim 1$, almost independently of parameters (in particular, $\sin\beta$ can be  small). This means that there is a strong tendency to produce significant dipole emission near the ISCO ($r\sim 1$).

\section{Jets}
We have so far shown that misaligned magnetic accretion can lead to dipole emission at the LT precession frequency. We will consider later the conditions needed for this to be strong, but first ask what the effect of significant dipole emission could be. The important point is that the emission frequency is always far below the plasma frequency of surrounding matter: from (\ref{LT}) we have
\begin{equation}
  \omega < \frac{2c}{R_g} = 4\times 10^4M_{10}^{-1}\, {\rm s^{-1}},
  \label{LT2}
\end{equation}
where $M_{10} = M/10M_\odot$, which compares with the plasma frequency
\begin{equation}
  \omega_{\rm pl} \simeq 9000N^{1/2} \sim 10^{14}M_{10}^{-2}\, {\rm s^{-1}},
  \label{pl}
\end {equation}
where $N$ is the electron number density in cm$^{-3}$, and we have inserted a typical midplane disc density at the second step. The effect is to absorb all of this extremely low--frequency emission and/or refract it into the regions of lowest density. In an accretion disc this is always away from the disc plane, which for strong dipole emission near the ISCO means towards the black hole spin axis, a near--vacuum because of the centrifugal barrier. (The emission pattern (\ref{dP/do}) is already a factor 2 stronger in these directions.) Evidently an accretion disc with strong dipole emission from the vicinity of the black hole is unstable to driving outflows in these directions, since once they appear the electron density remains lower there, acting as a stable escape route for further outflow. So in these states most of the primary dipole luminosity drives gas outflows confined close to the spin axis. There is an obvious analogy with radio pulsars, the other astrophysical objects where dipole emission is a major part of the energy output. Pulsars are observed to produce jets directed along the spin axis of the neutron star \citep{Markwardt:1995aa} about which the magnetic field rotates (rather than precessing, in the accretion disc case). We therefore suggest that misaligned accretion on to a black hole can similarly produce jets in suitable regimes, particularly if these are launched close to the ISCO. A promising feature of this idea is that dipole emission is initially coherent, offering the chance of driving nonthermal processes and so high brightness temperatures in the jets. This is already familiar, if not yet fully understood, for radio pulsars.

\section{Quasi--Periodic Oscillations}
The sharp peak of the dipole power output at the innermost radius where the accretion disc has a significant misalignment provides a natural origin for the quasi--periodic oscillations (QPOs) seen in X--rays from accreting systems. It has long been suspected that QPO frequencies correspond to LT precession near the inner edge of a misaligned disc in X--ray binaries and AGN \citep[e.g.][]{Ipser:1996aa,Stella:1999aa}. The problem has always been to specify a physical mechanism modulating the X--rays at these frequencies. Dipole emission explicitly produces instantaneous (i.e. not time--averaged) power with intensity modulated at LT frequencies, without needing any particular geometry. The sharp peak in dipole power means that effectively only the LT frequency from one radius appears. As this radius can itself vary, the modulation is quasiperiodic rather than strictly so. This same radius is also the site where any jet is launched, in agreement with frequent suggestions in the literature that the QPO is produced at the `base' of the jet.

Although the emitted power is proportional to $\omega^4$, it is unlikely that this translates to a relation between observed QPO strength and frequency, because the primary coherent radiation is always strongly absorbed and re--emitted at a much wider range of electromagnetic frequencies.

\section{A Minimum Condition for Jet Production}
We have argued above that significant dipole emission from misaligned accretion disc annuli is a promising way of driving jets.  As these are powered by the local accretion energy release it follows that they are likely to be stronger at smaller disc radii. The strongest jets must therefore correspond to cases where the dipole emission can remove a significant fraction of the accretion energy at the ISCO. Conversely, jets of this type are suppressed wherever the disc aligns completely with the black hole spin, so that $\beta = 0$, or the hole has zero spin ($a = 0$) -- the two types of stationary black hole state allowed by Hawking's theorem. 

So a minimum condition for jets to appear is that the dipole emission timescale, or equivalently the dipole alignment timescale $t_{\rm dip}$, should be shorter than the local timescale $t_2$ for viscous alignment \citep[see e.g.][]{King:2013ab,Nixon:2016aa}. For a disc annulus of radial width of order its scaleheight $H$ we use standard disc formulae \citep[e.g.][]{Frank:2002aa} to find its angular momentum $\Delta J$, and use the standard dipole alignment torque (\ref{Td}) (equivalently we can multiply the alignment time $t_h$ (eqn \ref{th}) by $\Delta J/J$, where $J = GM^2a/c$ is the hole angular momentum). This gives
\begin{equation}
t_{\rm dip} = t_h\frac{\Delta J}{J} =
\frac{3}{2a}\left(\frac{c_s}{v_A}\right)^2r^3\frac{GM}{c^3}
\label{tdip}
\end{equation}
and
\begin{equation}
  \label{tnu2}
  t_2 = 2\alpha\left(\frac{R}{H}\right)^2 r^{3/2} \frac{GM}{c^3},
\end{equation}
so that
\begin{equation}
\frac{t_{\rm dip}}{t_2} = \frac{3}{4a\alpha}\left(\frac{H}{R}\right)^2\left(\frac{c_s}{v_A}\right)^2r^{3/2}.
\label{t/t}
\end{equation}
Even at $r \sim 1$, where dipole emission is strongest, we see that for strong jets to appear (i.e. $t_{\rm dip}< t_2$) we require
\begin{equation}
\frac{H}{R} \la (a\alpha)^{1/2}\frac{v_A}{c_s} \la 0.1 - 0.37
\label{jet}
\end{equation}
where we have taken $\alpha \sim 0.1$, $v_A < c_s$ and $a = 0.1 - 1$ at the last step. This shows that strong jets are likely to be suppressed in discs with large aspect ratios $H/R$. In particular, significant radiation pressure ($H/R \sim 1$) is not promising for strong jet emission. This agrees well with inferences from X--ray binary state changes \citep[e.g.][]{Belloni:2010aa}, where jets appear in low states and are suppressed in high states. In the low--hard X--ray state, low--frequency QPOs usually interpreted as LT precession often appear, suggesting that $H/R$ cannot be very large, since a near--spherical flow cannot show significant precession (see Nixon \& Salvesen, 2014 for a discussion).

Of course a local (fixed $r$) analysis of this kind cannot give {\it sufficient} conditions for jet production -- for example, the disc may have already aligned with the spin plane before reaching small $r$. To find sufficient conditions we need to follow the alignment process globally, with simulations treating all disc radii simultaneously. We will attempt this in a future paper.

\section{Discussion}
This paper has suggested that misaligned disc accretion gives a physical origin for jets and quasiperiodic oscillations in accreting black hole systems. We have given simple criteria for such jets to appear. The driving process is the tendency towards axisymmetry through electromagnetic energy loss, which favours emission along the black hole spin axis provided that accretion is still misaligned near the inner disc edge. 

Although dipole emission requires the black hole to be spinning, its rotational energy nevertheless remains fixed, and the energy lost in jets is supplied purely by the accretion energy of the infalling disc annuli. These must contract, releasing gravitational binding energy. In this sense the jet luminosity removes a component of the accretion luminosity before the disc can radiate it.  This is completely distinct from the Blandford-Znajek (BZ) process \citep{Blandford:1977aa}, which explicitly taps the spin energy of the black hole. (This is also true of the fixed--field, aligning black hole spin case considered by \cite{King:1977aa} and discussed in the Introduction -- the work done by the torque ${\bf T}$ (eqn {\ref{T}) is formally identical with the BZ power up to an inclination factor $\sin^2\beta$.) A consequence of this is that the dipole emission process discussed here can launch jets from radii outside the black hole ergosphere, and in systems where the accretor is not a black hole (see below), but BZ cannot. The two processes differ further in that dipole emission is a purely vacuum electromagnetic process, whereas BZ requires the existence of a force--free magnetosphere. 

MHD and GRMHD formulations cannot directly treat dipole emission because of the neglect of the displacement current linking field accelerations to radiation. These simulations calculate the evolution of magnetic fields which move subluminally (with electric fields $E \ll B$). These are not radiation fields, but one could use the local magnetic field components and calculate the magnetic moments $\mu\sin\beta$ of disc annuli and so the implied dipole emission from (\ref{dP/do}, \ref{P}). This would allow a check of whether current simulations are self--consistent in neglecting dipole emission, and in particular, its back--reaction on the gas dynamics. The steep radial dependence (cf equation~\ref{Ld}) of dipole emission evidently requires very high numerical resolution near the inner disc edge here.

A more ambitious procedure would try to include dipole emission and the back--reaction torque self--consistently. But here one runs into the problem of how to treat the refraction and absorption of the coherent dipole emission at frequencies well below the plasma frequency (cf equations \ref{LT2}, \ref{pl}). Here the 1--fluid approximation is inadequate, and we need a plasma code, with a vast increase in computational complexity. This is not surprising given that jets can have high Lorentz factors and high brightness temperatures. The experience of trying to describe pulsar magnetospheres -- an inherently simpler (non-relativistic) problem, is not encouraging. The approach we have adopted here is to try to delineate the regimes where such plasma processes must occur, so identifying the likely production of jets.

The picture we have presented here agrees with a number of observed features. First, if jets result from misaligned accretion, they should appear more readily in AGN, where there is no natural plane for disc accretion, than in X--ray binaries, where discs always form in the orbital plane and make weaker misalignment natural. This appears to agree with observation. Further, we have shown that magnetic alignment is less effective in discs where radiation pressure is significant, so that jets should be weaker or absent in high accretion states, and relatively stronger in low states. Again this agrees with observation. 

In a future paper we shall investigate the detailed conditions for jets to arise through misaligned accretion in X--ray binaries. There are obvious hysteresis effects here if other mechanisms have already removed misalignment in more distant parts of the disc before magnetic effects can become important. This connects naturally with the explanation of X--ray binary state changes suggested by \citep{Nixon:2014aa}. Until this is done, we cannot judge whether jet power actually has the strong dependence on the black hole spin parameter $a$ suggested by (\ref{Ld}).

Any cogent model for jet production must explain not only why black hole systems often have jets and QPOs, but why almost all other accreting systems also do. The black--hole systems considered here use the LT precession of inclined orbits to drive magnetic dipole emission to do this. Systems with less compact but spinning accretors (neutron stars, white dwarfs and protostars) should drive (actually rather stronger) differential precession of inclined gas orbits (always with frequencies $\omega \sim R^{-3}$), simply because the accretors must have significant mass quadrupole moments. We therefore expect similar jets and QPOs in these systems too. Importantly, these non--relativistic systems may offer an easier route to understanding jet and QPO production than the full GR problem which black hole accretion presents.

\acknowledgments
We thank Jean--Pierre Lasota, Luciano Burderi, and the anonymous referee for helpful comments. CJN is supported by the Science and Technology Facilities Council (grant number ST/M005917/1). The Theoretical Astrophysics Group at the University of Leicester is supported by an STFC Consolidated Grant.



\end{document}